\documentclass[11pt]{evn}
\usepackage[T2A]{fontenc}
\usepackage{epsf}
\usepackage[russian,english]{babel}
\usepackage{natbib}
\usepackage{xcolor}
\usepackage{hyperref}
\hypersetup{
            colorlinks=true,
            linkcolor=blue,
            filecolor=black,
            citecolor=black,
            urlcolor=blue
           }
\newcommand{\degr}{{}^\circ}

\definecolor{Dred}{rgb}{0.422,0.0,0.0}
\definecolor{Dblue}{rgb}{0.0,0.0,0.422}
\definecolor{Dgreen}{rgb}{0.0,0.422,0.0}

\newcommand{\Grb}[1]{\textcolor{Dgreen}{\bf #1}}
\newcommand{\Blb}[1]{\textcolor{Dblue}{\bf #1}}

\newcommand{\web}[1]{\Blb{\hyperlink{#1}{#1}}}
\def\aj{AJ}%
\def\apjs{ApJS}%
\def\iss{{\noindent\mbox{\small The 13$^{th}$ EVN Symposium \& Users Meeting Proceedings, 2016
}}}

\long\def\maintitle#1{{\vskip 0mm \begin{center}\section*{#1}\end{center}\nopagebreak[4]}}

\long\def\author#1{{\begin{center}\normalsize{\bf#1}\end{center}\vskip-1em\index{#1}}\nopagebreak[4]}
\long\def\address#1{{\begin{center}\small\noindent#1\end{center}\vskip-8mm}\nopagebreak[4]}

\begin{document}
\iss

\maintitle{VLBA Calibrator Survey 9 (VCS--9)}

\author{Leonid Petrov}

\address{Astrogeo Center, USA}

\begin{abstract}
   The goals, current status, and preliminary results of the VLBA
Calibration Survey VCS--9 are discussed.
\end{abstract}
{\bf Keywords}: {VLBI, Astrometry.}


\section{Motivation}

Nowadays, more than 2/3 of VLBA observations are made in a phase-referencing 
mode. Planning such observations is facilitated tremendously owing to the 
existence of the phase calibrator list that is taken from the Radio 
Fundamental Catalogue (RFC)\footnote{Available at 
\web{http://astrogeo.org/rfc}}. Despite the fact that the total number of 
VLBI astrometric sources by 2016.10.01 reached 11,444 objects, it is not 
uncommon to hear a complaint that ``there is no good calibrator just near 
my target field''. For reliable phase-referencing, a calibrator should be 
close to the target, preferably within 1--2$\degr$. Since known calibrators 
are distributed unevenly over the sky, there are still areas where their 
density is not sufficient.

\begin{table}[h]
   \begin{center}
      \begin{tabular}{rrrr}
          \hline
          dist        &  all    & astr & cal    \\
          \hline
          $1.0^\circ$ &  51.5\% & 49.4\% & 39.9\% \\
          $1.5^\circ$ &  79.7\% & 77.8\% & 67.8\% \\
          $2.0^\circ$ &  93.8\% & 92.8\% & 86.3\% \\
          \hline
      \end{tabular}
   \end{center}
   \par\vspace{-2ex}\par
   \caption{\sf The probability to find a known compact radio source with
            positions known at mas level within a disk of the 
            specified radius in any at $\delta > -30^\circ$.
            Sources from category ``astr'' have position accuracy better
            than 5~mas. Sources from category ``cal'' have 
            position accuracy better 5~mas and correlated flux density 
            brighter 25~mJy at baselines longer 3000 km. The table is 
            valid for 2015.01.01.
           }
   \label{t:prob}
\end{table}

  To overcome this problem, a VLBA project of further densification of the
calibrator list for the area with $\delta > -40^\circ$ was launched.
The first original goal of the project was to extend the spacial coverage
of VLBI astrometric catalogue to that level that any field of view
of Pan-STARRS (disk of $1.5^\circ$ radius) and/or LSST (disk of $1.75^\circ$ 
radius) will be guaranteed to have at least one VLBI calibrator. This would
allow to improve position accuracy of these catalogues. With a release of 
the Gaia catalogue of 1.14 billion stars with a median position accuracy of
2.3~mas this objective became irrelevant. However, a new goal emerged with 
the release of Gaia catalogue: extending the list of Gaia/radio counterparts
and a search for sources with statistically significant radio/optic offsets.

  The second goal of the survey is to study the population of steep 
spectrum sources. Traditionally, flat-spectrum sources were scheduled for 
VLBI surveys. The pool of known flat spectrum sources brighter 50~mJy at 
4.8~GHz (single dish) at $\delta > -30^\circ$ without prior VLBI observations
had only 2284 objects at the beginning of the project and has already been 
depleted. Remaining sources either have spectrum steeper than $-0.5$
($S\sim f^{+\alpha}$) or have unknown spectrum. The population of steep 
spectrum sources is poorly studied due to a heavy selection bias in the past.

\section{Observations}

  The observing campaign started on 2015.08.07. The target list is 
combined from three sets:

\par\vspace{-1.2ex}\par
\begin{itemize}\itemsep0.4ex \parskip0.4ex
    \item all the sources from GB6 \citep{r:gb6} and PMN 
          \citep{r:pmn} catalogues at $\delta > -40\degr$ brighter 
          70~mJy at 4.85~GHz, except those observed in prior surveys, 
          known planetary nebulae and HII regions within $40''$;

    \item all the sources from the NVSS catalogue \citep{r:nvss}
          at $\delta > -40\degr$ and $|b|>5\degr$ brighter 200~mJy, 
          except those observed in prior surveys and present in 
          the first list;

    \item $\sim\! 400$ peculiar sources: objects with large position 
          offsets between radio and preliminary version of Pan-STARRS 
          catalogues, candidates for association with {\it Fermi},
          known sources with poor position accuracy, etc.
\end{itemize}

  Observations are scheduled in one scan of 60~s long at remote wings
of the C-band receiver: at [4.128, 4.608] and [7.392, 7.872]~MHz
simultaneously, single polarization, 2~bit sampling, aggregate  bit rate 
2~Gbps. The campaign was scheduled totally automatically: the array 
operator 1--40 hours in advance selects the start and stop dates of 
a segment and using Web interface executes a remote program that generates 
key-file with the schedule. The scheduling software assigns a score for 
every target source that is visible in accordance with slewing time and 
its priority. A source with the highest score is scheduled. The priorities
are assigned to increase chances of being observed for sources that are 
in the area of a low VLBI calibrator density, or just bright. 

  In addition to target objects, every hour 4 sources from a pool of known 
bright and compact objects are observed. These sources are selected in such 
a way that every station has two observations at elevations 
$[15\degr, 35\degr]$ and two observations at elevations $[45\degr, 90\degr]$.
The VCS9 observing campaign ran in the so-called ``fill-in'' mode: 
segments from 3.5 to 10 hours long are scheduled when other high priority 
VLBA programs could not observe due to bad weather or failures of 1 or 2 
stations. Unlike to traditional observing campaigns where the principal 
investigator determines which sources will be observed, only statistical 
criteria are specified for scheduling campaigns in the fill-in mode. These 
criteria, alongside with availability of time slots at given ranges of local 
siderial time determine the probability of that a given source will 
be observed.

\section{Preliminary results}

  The data were correlated with the DiFX correlator with spectral resolution 
62.5~KHz and accumulation period length 100~ms. This correlator setup allowed
us to search for fringes in a very wide window limited mainly by the primary 
beam. The detection limit drops a factor of 2 at $3'$ at 7.6~GHz and at $5'$ 
at 4.4~GHz. This setup generates ~100~Gb visibilities per 1 hour observing 
time, which is a substantial amount but still manageable. The advantages of 
using a wide fringe search window are 1)~usage of low accurate input 
catalogues (typical errors of GB6 are 20--$40''$) is possible; 2) a compact 
component in an extended source that can be located up to several arcminutes 
away of the peak at a low-resolution image can be found; 3) a second source 
or a second component can be searched for within several arcminutes of the 
main target. 

\begin{figure}[h]
  \centerline{\epsfclipon\epsfxsize=0.86\textwidth  
              \epsffile{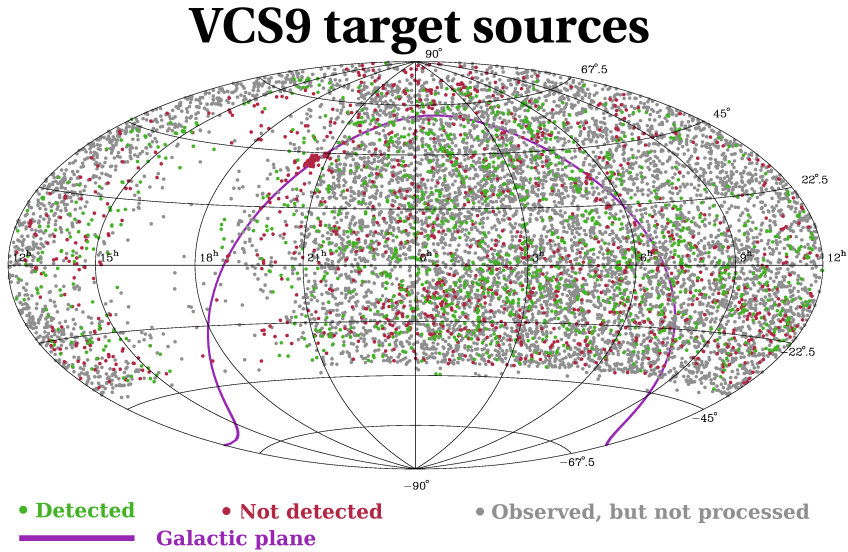}}
  \caption{\sf The distribution of observed target sources.}
  \par\vspace{-2ex}\par
  \label{f:vcs9_distr}
\end{figure}

  Figure~\ref{f:vcs9_distr} shows the distribution
of observed, detected, and non-detected sources. The zone within right 
ascensions 15--20~hours is under-observed because of a high demand 
for VLBA time by other high priority programs.

Table~\ref{t:vcs9_status} summarizes the status of the VCS9 campaign on 
2016.10.01. The output of the program is the catalogue of source positions 
and images of detected sources at 4.4 and 7.6 GHz. The preliminary results 
are accessible from the project web site \web{http://astrogeo.org/vcs9}. 
The median position accuracy of the catalogue is 1.2~mas, 90\% percentile 
is 7.2~mas. In total, 4973 images at both bands, including calibrators, 
were made available.

\begin{table}[h]
   \caption{\sf Status of the VCS9 project by 2016.10.01}
   \begin{tabular}{lr}
        Total time in hours:                                 &    536 \\
        The number of sources observed:                      &  \Blb{11012} \\
        The number of segments observed:                     &    99  \\
        The number of segments processed:                    &    36  \\
        The number of target sources processed:              &  2874  \\
        The number of target sources detected at any band:   &  \Grb{1583}  \\
        Detection rate:                                      &  \Grb{55\%} \\
   \end{tabular}
   \label{t:vcs9_status}
\end{table}

  Using preliminary results of VCS9 campaign, the distributions of sources 
over spectral index between 1.4 (NVSS) and 4.85 GHz (PMN, GB6) catalogue were 
computed. The main result can be formulated as ``a rule of one third'': one
third of all detected sources have steep spectra and one third of non-detected
sources have flat spectra. No attempt was made to evaluate completeness of 
the sample. This will be done later after completion of data analysis. But 
even incomplete results show clearly that the share of compact objects within
steep spectrum sources is much bigger than it was tacitly assumed before.

\begin{figure}[h]
  \epsfclipon\epsfxsize=0.48\textwidth  
             \epsffile{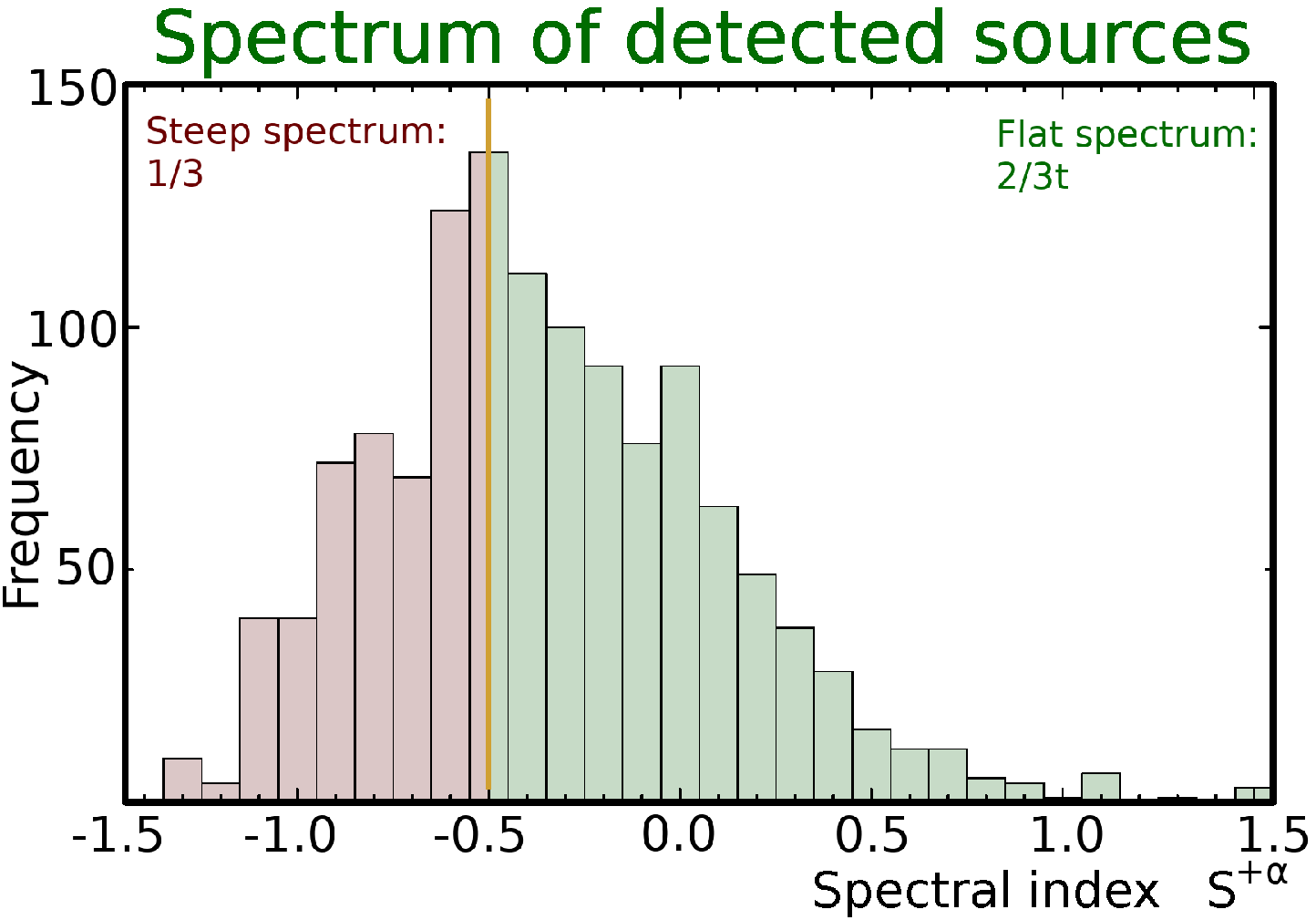}
             \hspace{0.03\textwidth}
  \epsfclipon\epsfxsize=0.48\textwidth  
             \epsffile{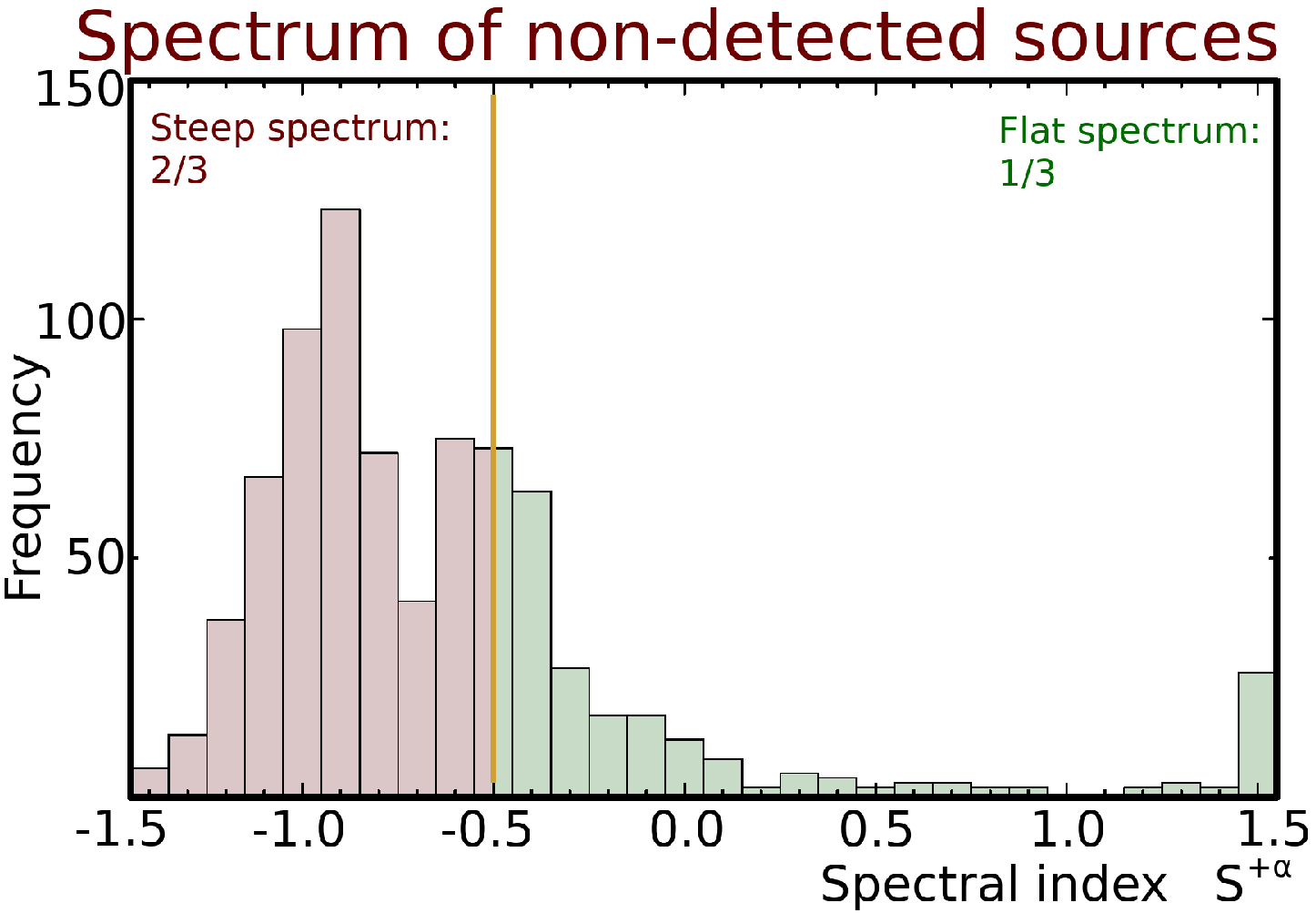}
  \par\vspace{-2ex}\par
  \caption{\sf The distribution of spectral indices for detected ({\it Left}) and
           not detected ({\it Right}) sources.}
  \label{f:det_ndt}
\end{figure}

\par\vspace{-1ex}\par


\begin{thebibliography}{3}\setlength{\itemsep}{0.2ex}
   \small
   \bibitem[Condon et~al.(1998)]{r:nvss}
           Condon,~J.~J. et al., \aj, 115, 1693

   \bibitem[Gregory et~al.(1996)]{r:gb6}
            Gregory,~P.~C., et al., 1996, \apjs, 103, 427

   \bibitem[Wright et~al.(1996)]{r:pmn}
            Wright,~A.~E., et al. 1996, \apjs, 103, 145

\end{thebibliography}
\end{document}